\documentclass[11pt]{article}
\usepackage{amssymb}
\usepackage{amsmath}
\usepackage{amstext}
\usepackage{graphicx,epsfig}
\usepackage{epsfig}
\usepackage{verbatim} 
\usepackage{fancybox}
\usepackage{color}
\usepackage{ulem}
\usepackage{enumitem}
\usepackage{subfigure}
\usepackage{bbm}
\usepackage{parskip}
\usepackage{dsfont}
\usepackage[numbers,sort&compress]{natbib}

\newcommand{\Comment}[1]{{}}
\definecolor{darkblue}{rgb}{0.15,0.35,0.55}
\definecolor{reddish}{rgb}{0.65, 0.2, 0.2}
\usepackage[linktocpage=true]{hyperref}
\hypersetup{
colorlinks=true,
citecolor=darkblue,
linkcolor=reddish,
urlcolor=darkblue,
pdfauthor={},
pdftitle={},
pdfsubject={}
}

\newcommand{\be}{\begin{equation}}
\newcommand{\ee}{\end{equation}}
\newcommand{\bea}{\begin{eqnarray}}
\newcommand{\eea}{\end{eqnarray}}
\newcommand{\beas}{\begin{eqnarray*}}
\newcommand{\eeas}{\end{eqnarray*}}
\newcommand{\nn}{\nonumber}

\def\({\left(}
\def\){\right)}

\newcommand{\Tr}{\text{Tr}}

\newcommand{\la}{\langle}
\newcommand{\ra}{\rangle}

\def\gsim{ \lower .75ex \hbox{$\sim$} \llap{\raise .27ex \hbox{$>$}} }
\def\lsim{ \lower .75ex \hbox{$\sim$} \llap{\raise .27ex \hbox{$<$}} }

\DeclareMathOperator{\SU}{SU}
\DeclareMathOperator{\Orth}{so}
\def\sech{\mathop{\rm sech}\nolimits}
\def\csch{\mathop{\rm csch}\nolimits}

\def\Tr{\mathop{\rm Tr}\nolimits}

\title{}
\author{}

\numberwithin{equation}{section}

\begin{document}
%
~
\vspace{2truecm}
\begin{center}
{\LARGE \bf{Holography for a Non-Inflationary Early Universe}}
\end{center} 

\vspace{1.3truecm}
\thispagestyle{empty}
\centerline{\Large Kurt Hinterbichler,${}^{\rm a}$ James Stokes ${}^{\rm b}$ and Mark Trodden${}^{\rm b}$}
\vspace{.7cm}

\centerline{\it ${}^{\rm b}$Perimeter Institute for Theoretical Physics,}
\centerline{\it  31 Caroline St. N, Waterloo, Ontario, Canada, N2L 2Y5}

\vspace{.3cm}

\centerline{\it$^{\rm b}$Center for Particle Cosmology, Department of Physics and Astronomy,}
\centerline{\it University of Pennsylvania, Philadelphia, PA 19104, USA}

\vspace{.4cm}

\begin{abstract}
\vspace{.03cm}
\noindent
We construct a gravitational dual of the pseudo-conformal universe, a proposed alternative to inflation in which a conformal field theory in nearly flat space develops a time dependent vacuum expectation value. Constructing this dual amounts to finding five-dimensional domain-wall spacetimes with anti-de Sitter asymptotics, for which the wall has the symmetries of four-dimensional de Sitter space. This holographically realizes the characteristic symmetry breaking pattern $\Orth(2,4) \to \Orth(1,4)$ of the pseudo-conformal universe.  We present an explicit example with a massless scalar field, using holographic renormalization to obtain general expressions for the renormalized scalar and stress-tensor one-point functions.  We discuss the relationship between these solutions and those of four-dimensional holographic defect conformal field theories which break $\Orth(2,4) \to \Orth(2,3)$.
\end{abstract}

\newpage


\section{Introduction}
In the most common examples of the AdS/CFT correspondence, the boundary field theories are $\SU(N)$ gauge theories and the bulk gravitational theories are string theories which reduce to Einstein gravity in an appropriate large-$N$ limit~\cite{Maldacena:1997re}. Indeed, as previously anticipated~\cite{Polyakov:1997tj}, 4D gauge theories in flat space should admit an exact description in terms of string theories on 5D backgrounds with a curved fifth dimension $\rho$
\begin{equation}\label{e:warped}
	ds^2 = d\rho^2 + a(\rho)^2 \eta_{\mu\nu}dx^\mu dx^\nu \ ,
\end{equation}
where $\eta_{\mu\nu}$ is the Poincar\'{e} invariant metric of the 4D gauge theory and $\rho$ is related to the Liouville (longitudinal) mode which arises in the quantization of the non-critical string. The gauge theory must be located at a position $\rho_\ast$ such that the warp factor either vanishes or becomes infinite. If the gauge theory is a conformal field theory (CFT) then conformal invariance fixes $a(\rho) \sim e^{\rho/R}$, corresponding to $\mathrm{AdS}_5$ with radius $R$, and it is convenient to choose $\rho_\ast = \infty$ which corresponds to placing the CFT at the boundary of $\mathrm{AdS}$. The AdS/CFT correspondence can thus be viewed as a particular example of a more general gauge-string duality. Wick rotating this picture, $a(\rho)$ is essentially a scale factor, and the conformal limit is de Sitter space \cite{Strominger:2001pn,Maldacena:2002vr}.  This observation has been used to provide an alternative way to look at the behavior of inflationary spacetimes in the early universe, in which a Euclidean CFT at the asymptotic future is dual to an inflationary spacetime in the bulk \cite{McFadden:2009fg,McFadden:2010na,Bzowski:2012ih}.

This raises the question of whether it is possible to holographically realize other, non-inflationary proposals for the physics of the early universe.
The pseudo-conformal universe \cite{Rubakov:2009np,Creminelli:2010ba,Hinterbichler:2011qk,Hinterbichler:2012mv} is an early universe scenario whose characteristic feature is spontaneous symmetry breaking of the conformal group in 4D to a subalgebra which is isomorphic to the algebra of de Sitter isometries.  The idea is to postulate that the early universe is described by a CFT containing a set of scalar primary operators $\mathcal{O}_I$ with conformal dimensions $\Delta_I$. The CFT is assumed to exist in a state which spontaneously breaks the $\Orth(2,4)$ conformal algebra to an $\Orth(1,4)$ de Sitter subalgebra, which happens when the operators develop time-dependent vacuum expectation values of the form
\begin{equation}\label{e:vev}
	\langle \mathcal{O}_I \rangle \sim \frac{1}{(-t)^{\Delta_I}}.
\end{equation} 
The time $t$ runs from $-\infty$ to $0$, and the scenario breaks down at late times near $t=0$, where a re-heating transition to a traditional radiation dominated expansion must take place.   
This symmetry breaking pattern ensures that spectator fields of vanishing conformal dimension will automatically acquire scale-invariant spectra \cite{Hinterbichler:2011qk}, which under favorable conditions can be transferred to become the scale invariant curvature perturbations we see today \cite{Wang:2012bq}.

During the pseudo-conformal phase, spacetime is approximately flat, in contrast to the  exponential expansion of inflation.  As a consequence, gravity wave production in the pseudo-conformal scenario is exponentially suppressed.  Thus, an observation of a large primordial tensor to scalar ratio (for example the interpretation of the result reported by BICEP2 \cite{Ade:2014xna}) would provide a crucial test of the pseudo-conformal proposal.

In this paper, we will be interested in constructing a bulk dual to a CFT in the pseudo-conformal phase.  In the case of dS/CFT and duals to inflation, the physics of interest is in the bulk and the dual is the boundary CFT.  Here, by contrast, the physics of interest is the non-gravitational CFT on the boundary, and the dual is the bulk gravitational theory.

If a CFT possesses a gravitational dual, then the conformal vacuum corresponds to empty $\mathrm{AdS}$, and other states of the CFT Hilbert space correspond to activating configurations of fields in the bulk. These field configurations can break the $\Orth(2,4)$ isometry group of $\mathrm{AdS}_5$ to a subgroup, which in turn breaks the isometry group of the spacetime through gravitational backreaction. At large values of the radial coordinate $\rho$, however, the warp factor should approach $\sim e^{\rho/R}$, corresponding to restoration of the full conformal group at high energies. In general the broken/unbroken isometries of the asymptotically anti-de Sitter background map to the corresponding broken/unbroken conformal generators of the field theory. It follows that to implement the pseudo-conformal mechanism in AdS/CFT, we need a spacetime with the isometries of $\Orth(1,4)$, which are enhanced to $\Orth(2,4)$ at the boundary. Geometrically, this corresponds to a domain-wall spacetime asymptoting to anti-de Sitter space, where the domain wall is foliated by four-dimensional de Sitter slices.  Since the isometry group of $\mathrm{dS}_4$ is $\Orth(1,4)$, this realizes the required breaking pattern. In the limit in which backreaction is ignored, and near the boundary, this should revert to $\mathrm{AdS}_5$ in the de Sitter foliation.

In Section \ref{probesections} we will first consider the simpler case of a probe scalar, ignoring backreaction, in which the background profile for the scalar should be preserved by a $\Orth(1,4)$ subgroup of $\Orth(2,4)$. We find exact solutions of the wave equation for any value of the mass of the scalar.  We then generalize in Section \ref{backreacsect} to the fully interacting Einstein-scalar theory and obtain the background equations of motion for domain-wall spacetimes which have the appropriate symmetries, providing an explicit solution for the case of a massless scalar.   We determine the VEV of the operators dual to the metric and scalar field and show that they have the correct form \eqref{e:vev} to break the conformal group of the boundary field theory to a de Sitter subalgebra.  We review the appropriate holographic renormalization machinery in Appendix \ref{hrenormappe}, in which we calculate the exact renormalized one-point functions for arbitrary source configurations.

\section{Probe Scalar Limit\label{probesections}}

As a warmup for the full back-reacted problem, we first consider a probe scalar on $\mathrm{AdS}_{5}$. 
The coordinates adapted for working with a dual Minkowski CFT are those of the Poincar\'{e} patch, in which $\mathrm{AdS}_{5}$ is foliated by Minkowski slices parametrized by $x^\mu=(t,x^i)$,
\begin{equation}\label{e:planarAdS}
	ds^2 = \frac{1}{z^2}\left(dz^2 +\eta_{\mu\nu}dx^\mu dx^\nu\right) \ .
\end{equation}
Here the radial coordinate is $z$, which ranges over $(0,\infty)$, with $0$ the boundary and $\infty$ deep in the bulk.  We have set the $\mathrm{AdS}$ radius to unity.
The Killing vectors for $\mathrm{AdS}_5$ in these coordinates are
\begin{align}
	P_\mu
		& = -\partial_\mu ,\\
	L_{\mu\nu}
		& = x_\mu \partial_\nu - x_\nu \partial_\mu, \\
	K_\mu
		& = x^2 \partial_\mu - 2x_\mu x^\nu \partial_\nu + z^2 \partial_\mu - 2x_\mu z \partial_z ,\\
	D
		& = -x^\mu \partial_\mu - z\partial_z \ .
\end{align}
The first two sets of Killing vectors are the generators of the 4D Poincar\'{e} subalgebra ${\rm iso}(1,3)$ preserved by constant $z$ slices.  Going to the boundary at $z = 0$, the Killing vectors become the generators of the 4D conformal group $\Orth(2,4)$, which is the statement that the isometries of anti-de Sitter act on the boundary as conformal transformations.

A configuration of a bulk scalar $\phi$ of mass $m$ will have an expansion near the boundary of the form
\be \phi(x,z) =  z^{\Delta_-}\left[\phi_0(x)+{\cal O}(z^2)\right]+z^{\Delta_+}\left[\psi_0(x)+{\cal O}(z^2)\right],\label{FGprobes}\ee
where 
\be
\Delta_\pm = 2 \pm \sqrt{4 + m^2}.
\ee
(There can be additional logarithmic terms if $\sqrt{4 + m^2}$ is an integer.)  The coefficient $\phi_0(x)$ is a source term in the lagrangian of the CFT which sources an operator $\cal O$ of dimension $\Delta_+$, and
the coefficient $\psi_0(x)$ determines the vacuum expectation value (VEV) of this operator \cite{Klebanov:1999tb},
\be \la {\cal O}\ra=(2\Delta_+-4)\psi_0. \ee
We are interested in the case in which there is a VEV of the form \eqref{e:vev} in the absence of sources, so that the symmetries are spontaneously broken by the VEV rather than explicitly broken by a source.  Thus we seek configurations for which $\psi_0\sim {1/ (-t)^{\Delta_+}}$ and $\phi_0=0$.

The field profile we want must preserve the $D$, $P_i$, $L_{ij}$ and $K_i$ conformal generators, which form the unbroken de Sitter $\Orth(1,4)$ subalgebra of interest \cite{Hinterbichler:2011qk}. We seek the most general bulk scalar field configuration which preserves these symmetries. Preservation of the spatial momentum and rotations, $P_i$ and $L_{ij}$, implies that the scalar depends only on $t$ and $z$,
\begin{equation}
	\phi = \phi(z,t) \ .
\end{equation}
Demanding invariance under $D = -z \partial_z - x^\mu \partial_\mu$, gives
\begin{equation}
	 z \partial_z \phi + t \partial_t \phi=0 \ ,
\end{equation}
which implies that the field is a function only of the ratio $z/t$,
\begin{equation}
	\phi = \phi(z/t) \ .\label{profile1}
\end{equation}
This is now automatically invariant under the spatial special conformal generators, $K_i$,
\begin{equation}
	K_i \phi = -2x^i (z \partial_z + t \partial_t)\phi = 0 \ .
\end{equation}
Thus, a profile $\phi(z/t)$ is the most general one which preserves the required $\Orth(1,4)$ symmetries. 

\begin{figure}[h!]
\centering
\label{coordinates}
\includegraphics[width=8cm]{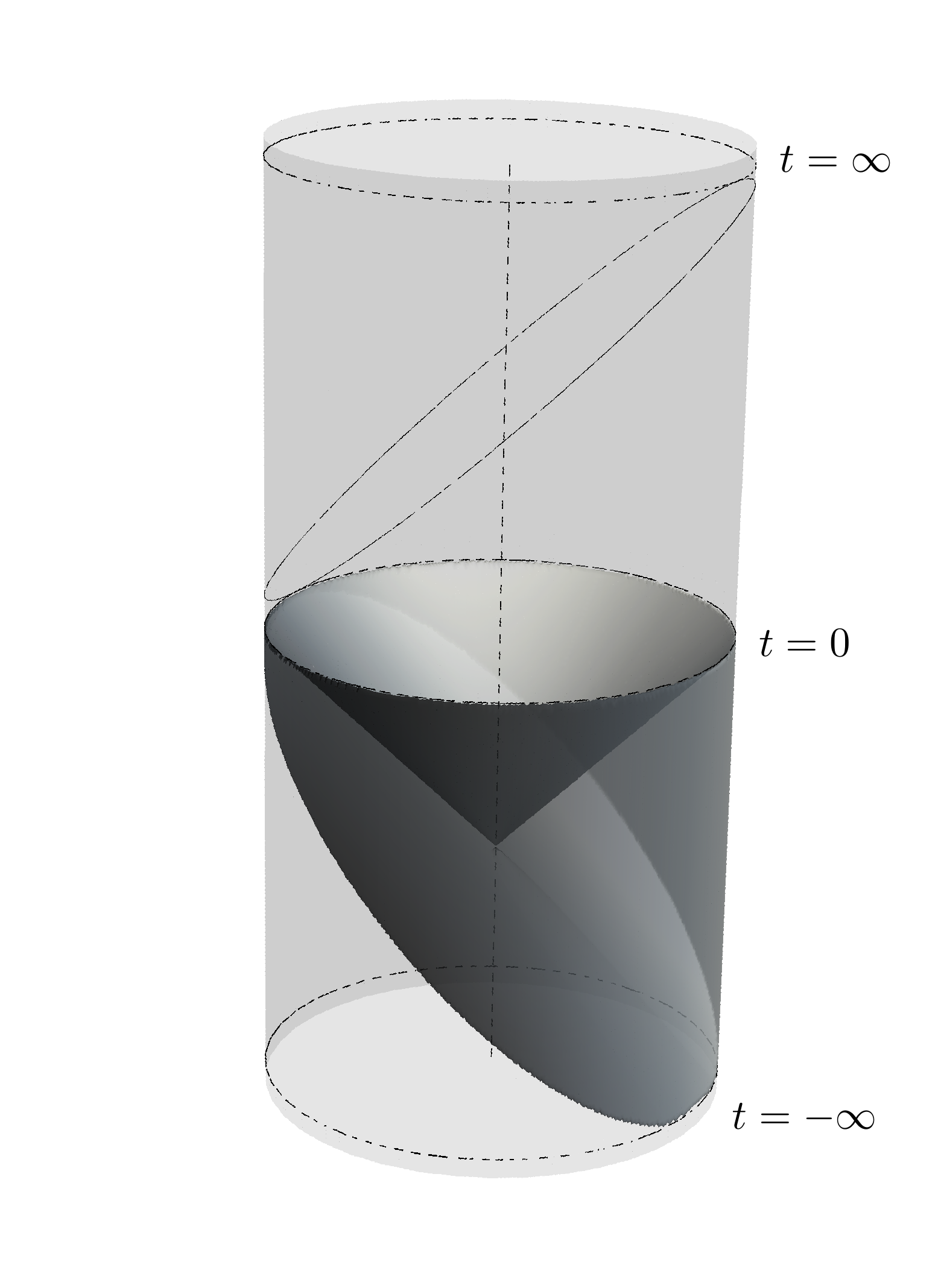}\ \ \ \includegraphics[width=8cm]{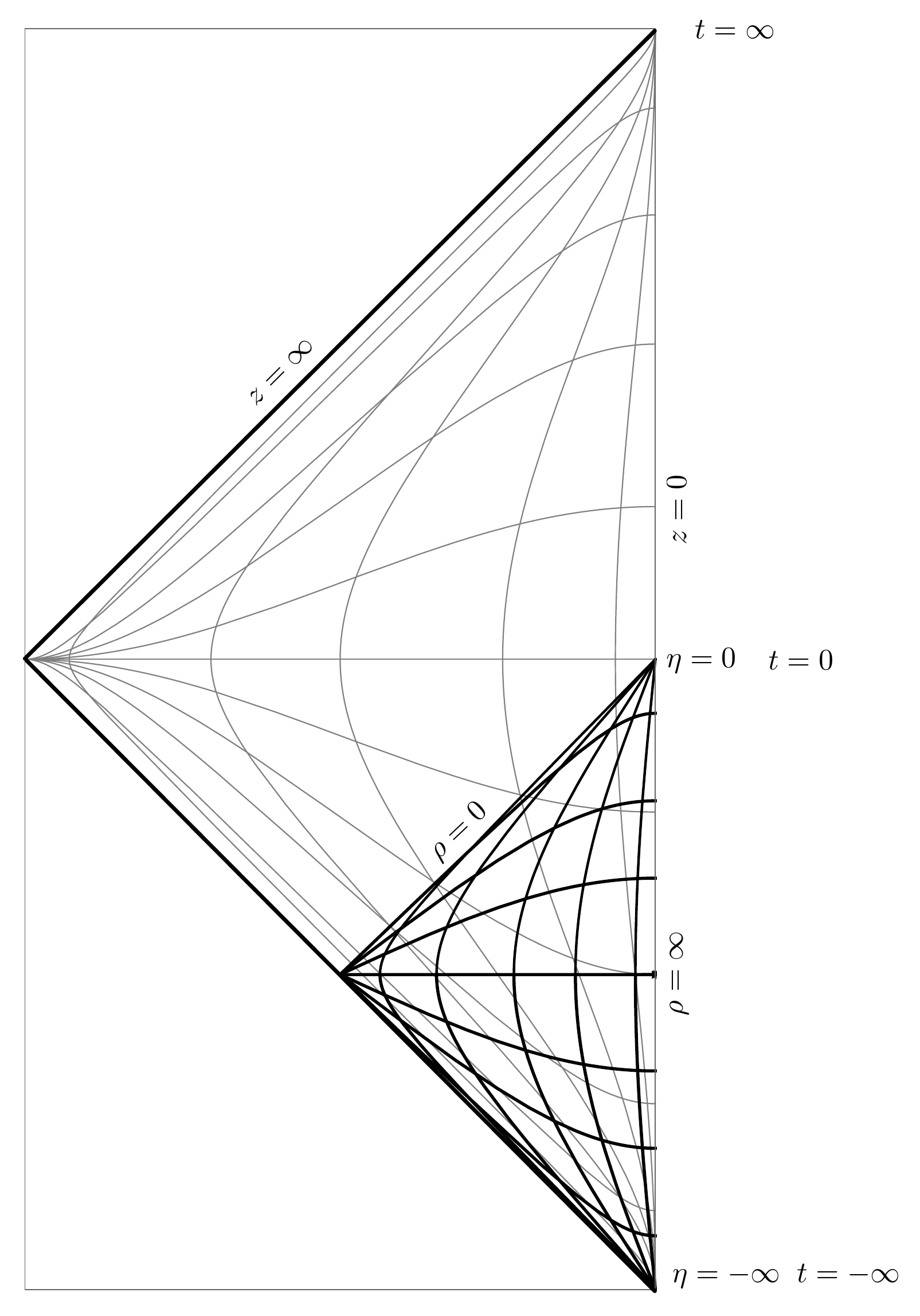}
\caption{Penrose diagrams showing the Poincar\'{e} coordinates and de Sitter slice coordinates.  The left-hand figure shows the global AdS cylinder; the de Sitter slice coordinate region is bounded from above by the lightcone which emanates downward from $t=0$, and is bounded from below by the slanted ellipse, which also marks the lower boundary of the Poincare patch (the upper slanted ellipse shown in outline form is the upper boundary of the Poincar\'{e} patch).  The right-hand figure shows a two dimensional slice down the axis of the AdS cylinder; thin lines are Poincar\'{e} lines of constant $z$ and $t$, thick lines de Sitter slice lines of constant $\rho$ and $\eta$.  }
\end{figure}

It will be useful to work in coordinates adapted to the unbroken $\Orth(1,4)$ symmetries.  
Define a radial coordinate $\rho\in (0,\infty)$ and a time coordinate $\eta \in (-\infty,0)$ by the relations
\bea && t=\eta \coth\rho,\ \ \ z=(-\eta)\csch\rho,\nn\\
&&  \rho=\cosh^{-1}\left[{(-t)\over z}\right],\ \ \ \ \eta=-\sqrt{t^2-z^2}.
\eea
In these coordinates, the metric becomes
\begin{equation}\label{e:dSslicing}
	ds^2 = d\rho^2 + \sinh^2\rho\left[\frac{-d\eta^2 + d\vec{x}^2}{\eta^2}\right] \ ,
\end{equation}
which we recognize as the foliation of $\mathrm{AdS}_5$ by inflationary patch $\mathrm{dS}_4$ slices.   These coordinates cover the region $t<0$, $(-t)>z$, which is the bulk causal diamond associated with the time interval $t=(-\infty,0)$.  This is the region we expect to be dual to the pseudo-conformal scenario\footnote{Note that this region does not satisfy the criterion proposed in \cite{Bousso:2012mh}, which suggests that a full duality may require non-local operators in an essential way.  See also \cite{Bousso:2012sj,Czech:2012bh,Hubeny:2012wa,Leichenauer:2013kaa}.}.  The boundary is approached as $\rho\rightarrow\infty$, and the line $(-t)=z$ is approached as $\rho\rightarrow 0$.  Near the boundary, the coordinate $\eta$ corresponds with $t$, and $\rho$ approaches
$ e^{\rho} \simeq 2{(-t)\over z}$.
The region and coordinates are illustrated in Figure \ref{coordinates}.

We now consider the scalar wave equation in these coordinates. The equation of motion for a minimally coupled real scalar field of mass $m$ on curved space is
\begin{equation}
	(\square^{(5)} - m^2)\phi =0\ .
\end{equation}
In the de Sitter adapted coordinates, a scalar configuration with the desired profile \eqref{profile1} is one which depends only on the $\rho$ coordinate. The wave equation then reduces to
\begin{equation}
	\phi''(\rho) + 4 \coth \rho \, \phi'(\rho) - m^2 \phi(\rho)=0 \ .
\end{equation}
The general solution is
\bea
	\phi(\rho)
		&=& C_+\frac{e^{-\sqrt{4+m^2} \rho} \left(\sqrt{4+m^2}+\coth \rho\right) \csch^2\rho}{\sqrt{4+m^2}} \nn \\
		&&\ \ \ \ \ \ \ \ \ \ + C_-\frac{e^{\sqrt{4+m^2} \rho} \left(\sqrt{4+m^2}-\coth \rho\right) \csch^2 \rho}{3+m^2},\ \ m^2 \neq -3,-4 \ , \nn\\
		\phi(\rho) &=& C_+ \csch^3 \rho +C_- \csch^3 \rho (\sinh 2\rho-2\rho) \ ,\ \ \ m^2=-3\, , \nn\\
		\phi(\rho) &=& C_+ \csch^2\rho (\rho\coth \rho - 1) + C_- \csch^2 \rho \coth \rho \ , \ \ \ m^2=-4\, , \label{gensols}
\eea
where $C_\pm$ are the two integration constants of the second order wave equation.
The mass must satisfy the Breitenlohner-Freedman stability bound $m^2 \geq -4$ \cite{Breitenlohner:1982bm}.  The mass range $m^2 > 0$ corresponds to irrelevant operators $\Delta_+>4$, $-4 \leq m^2 < 0$ corresponds to relevant operators $\Delta_+<4$, and $m^2=0$ to marginal operators $\Delta_+=4$.

At large $\rho$ (the boundary) the solutions \eqref{gensols} behave as
\begin{equation}
	\phi(\rho) \simeq C_+ e^{-\Delta_+\rho}\left[1+{\cal O}(e^{-2\rho})\right] + C_- e^{-\Delta_-\rho}\left[1+{\cal O}(e^{-2\rho})\right] \ ,
\end{equation}
where $\Delta_\pm = 2 \pm \sqrt{4 + m^2}$ and we have absorbed unimportant constants into $C_+,C_-$. (There are also terms proportional to $\rho$ and mixing between the two coefficients in the cases $m^2 = -3,-4$.)  Changing back to the Poincar\'{e} coordinates $z,t$ and using the asymptotic relation $e^{\rho} \sim {(-t)\over z}$, we have
\be \phi(\rho) \simeq C_+ z^{\Delta_+}\left[{1\over (-t)^{\Delta_+}}+{\cal O}(z^2)\right] + C_- z^{\Delta_-}\left[{1\over (-t)^{\Delta_-}}+{\cal O}({z^2})\right] \ .
\ee
(We have again absorbed unimportant constants into $C_\pm$, and there are also terms logarithmic in $z$ in the cases $m^2 = -3,-4$).  Comparing with \eqref{FGprobes} we see that this configuration has $\phi_0\sim {C_-\over (-t)^{\Delta_-}}$, $\psi_0\sim{C_+\over (-t)^{\Delta_+}}$.  Since we are interested in spontaneous breaking for which there is no source, $\phi_0=0$, we must fix $C_-=0$, after which we obtain a spontaneously generated VEV proportional to $\psi_0$,
\be \la {\cal O}\ra \sim {C_+\over (-t)^{\Delta_+}}.\ee

We thus have configurations which correctly reproduce the time-dependent vacuum expectation value required to break $\Orth(2,4) \to \Orth(1,4)$, where $\Delta_+$ is the scaling dimension of the dual operator, defined over the causal diamond of the region $t\in(-\infty,0)$ of the Poincar\'{e} patch. This is precisely what is required to realize the pseudo-conformal mechanism. 

A feature generic to these solutions is that they diverge as $\sim 1/\rho^3$ as $\rho \rightarrow 0$, which means the scalar is blowing up and back-reaction is becoming important as we approach the line $(-t)=z$ in the Poincar\'{e} patch.  This is dual to the approach to $t\to 0$ in the boundary, where the pseudo-conformal phase ends and additional physics must kick in to reheat to a traditional radiation dominated universe.  We now turn to the fully back-reacted case with dynamical bulk gravity.

\section{Including Back-Reaction\label{backreacsect}}

We now consider a scalar field minimally coupled to Einstein gravity
\begin{equation}\label{e:effectiveaction}
	S = \int d^5 x \sqrt{-G} \left[\frac{1}{2}R[G]+6-\frac{1}{2}G^{MN}\partial_M \phi\partial_N\phi - V(\phi)\right] - \int d^4 x \sqrt{-g} K \ .
\end{equation}
$G_{MN}$ is the bulk metric, and we have a Gibbons-Hawking term on the boundary \cite{Gibbons:1976ue,York:1972sj} which depends on the boundary metric and extrinsic curvature.  We have set the radius of $\mathrm{AdS}_5$ to unity and absorbed the overall factors of the Planck mass into the definition of the action.
We seek a metric which respects the unbroken $\mathrm{dS}_4$ isometries and approaches $\mathrm{AdS}_5$ at the boundary. The appropriate ansatz is thus a domain wall spacetime in which the wall is foliated by $\mathrm{dS}_4$ slices,
\begin{equation}\label{mansatzdw}
	ds^2 = d\rho^2 + e^{2A(\rho)} \left[\frac{-d\eta^2 + d\vec{x}^2}{\eta^2}\right] \ , \quad \phi = \phi(\rho) \ .
\end{equation}
We are interested in asymptotically $\mathrm{AdS}_5$ solutions, for which the scale factor $A(\rho) \to \rho+const.$ as $\rho \to\infty$.  We demand that the scalar potential has an extremum at $\phi = 0$ with the value $V(0)=0$ into which the scalar flows as $\rho\to \infty$,
\begin{equation}
	V =  \frac{1}{2}m^2 \phi^2 +\mathcal{O}(\phi^3) \ .
\end{equation}
The independent equations of motion for the background fields are\footnote{We note that these are the same background equations for a spherically symmetric Euclidean domain wall,
\begin{equation}
	ds^2 = d\rho^2 + a(\rho)^2 (d\theta^2 + \sin^2\theta d\Omega_3^2) \ .
\end{equation}
This can be seen by analytically continuing the angular coordinates as $\theta = i t + \pi / 2$. The $S^4$ metric continues to the global $\mathrm{dS}_4$ metric and we obtain a domain wall foliated by global $\mathrm{dS}_4$ slices
\begin{equation}
	ds^2 \to d\rho^2 + a(\rho)^2(- dt^2 + \cosh^2 t \, d\Omega_3^2) \ .
\end{equation}
Under the analytic continuation $\rho = i t$, $\theta = i\lambda$, $\hat{a}(t) = - i a(it)$, the spherically symmetric Euclidean domain wall maps to a FLRW spacetime with hyperbolic spatial sections $ds^2 = -dt^2 + \hat{a}(t)^2(d\lambda^2 + \sinh^2\lambda d\Omega_3^2)$. This analytic continuation was used in \cite{Hertog:2005hu,Hertog:2004rz,Craps:2007ch} to holographically study crunching $\mathrm{AdS}$ cosmologies.}

\begin{align}
	\phi''(\rho) + 4A'(\rho)\phi'(\rho)
		& = \frac{\partial V}{\partial \phi}, \label{e:scalareom} \\
	6A'(\rho)^2 - 6e^{-2A(\rho)}-6
		& = \frac{1}{2} \phi'(\rho)^2 - V \label{e:constraint} \ .
\end{align}
When $\phi = 0$, we find that the equations are solved by $A(\rho)=\log(\sinh \rho)$, which is empty $\mathrm{AdS}_5$ corresponding to the conformal vacuum of the dual field theory.

A simple solution of the second-order system can be found by choosing a vanishing potential for the scalar, $V(\phi) =0$.  This choice corresponds to a particular truncation of type IIB supergravity, as we review in Appendix \ref{sugraappen}.  The scalar is the string theory dilaton, and the dual operator is closely related to the $\mathcal{N} = 4$ SYM Lagrangian \cite{Clark:2004sb}, which is marginal, $\Delta=4$. 

Now that we have chosen $V=0$, we may integrate \eqref{e:scalareom} once to obtain
\begin{equation}\label{e:scalar}
	\phi'(\rho) = c e^{-4A(\rho)} \ ,
\end{equation}
with integration constant $c$.  Substituting this into \eqref{e:constraint} we obtain
\begin{equation}\label{e:constraintnew}
	A'(\rho)^2 = 1 + e^{-2A(\rho)} + b e^{-8A(\rho)} \ ,
\end{equation}
where $b \equiv c^2/12\geq 0$. Defining a new coordinate,
\be u = e^{-2A(\rho)}\label{ucoorddef} \ ,
\ee
we obtain, with the use of \eqref{e:constraintnew}, the metric \eqref{mansatzdw} in the form
\begin{equation}\label{e:janus}
	ds^2 = \frac{du^2}{4u^2(1 + u + b u^4)} + \frac{1}{u} \left[\frac{-d\eta^2+d\vec{x}^2}{\eta^2}\right] \ .
\end{equation}
This metric is $\mathrm{AdS}_5$ for $b = 0$ and for $b \neq 0$ it approaches $\mathrm{AdS}_5$ near the boundary at $u = 0$. 
We have arrived at a general solution for the metric with a single integration constant.  This is one of the three integration constants expected in the general solution of the original system \eqref{e:scalareom}, \eqref{e:constraint}.  Of the other two, one is expressed as an arbitrary shift on the scalar (since the scalar only appears with derivatives) and the other can be absorbed into the scale of the de Sitter slices (which can be fixed to unity by demanding $A(\rho)\to \rho-\ln 2$ at infinity).
The equation for the scalar \eqref{e:scalar} expressed in terms of the variable $u$ is
\begin{equation}
	\frac{d{\phi}}{du} = -\frac{c}{2} \frac{u}{\sqrt{1+u + b u^4}} \ .
\end{equation}

\begin{figure}[h]
\centering
\label{solutionplot}
\includegraphics[width=12cm]{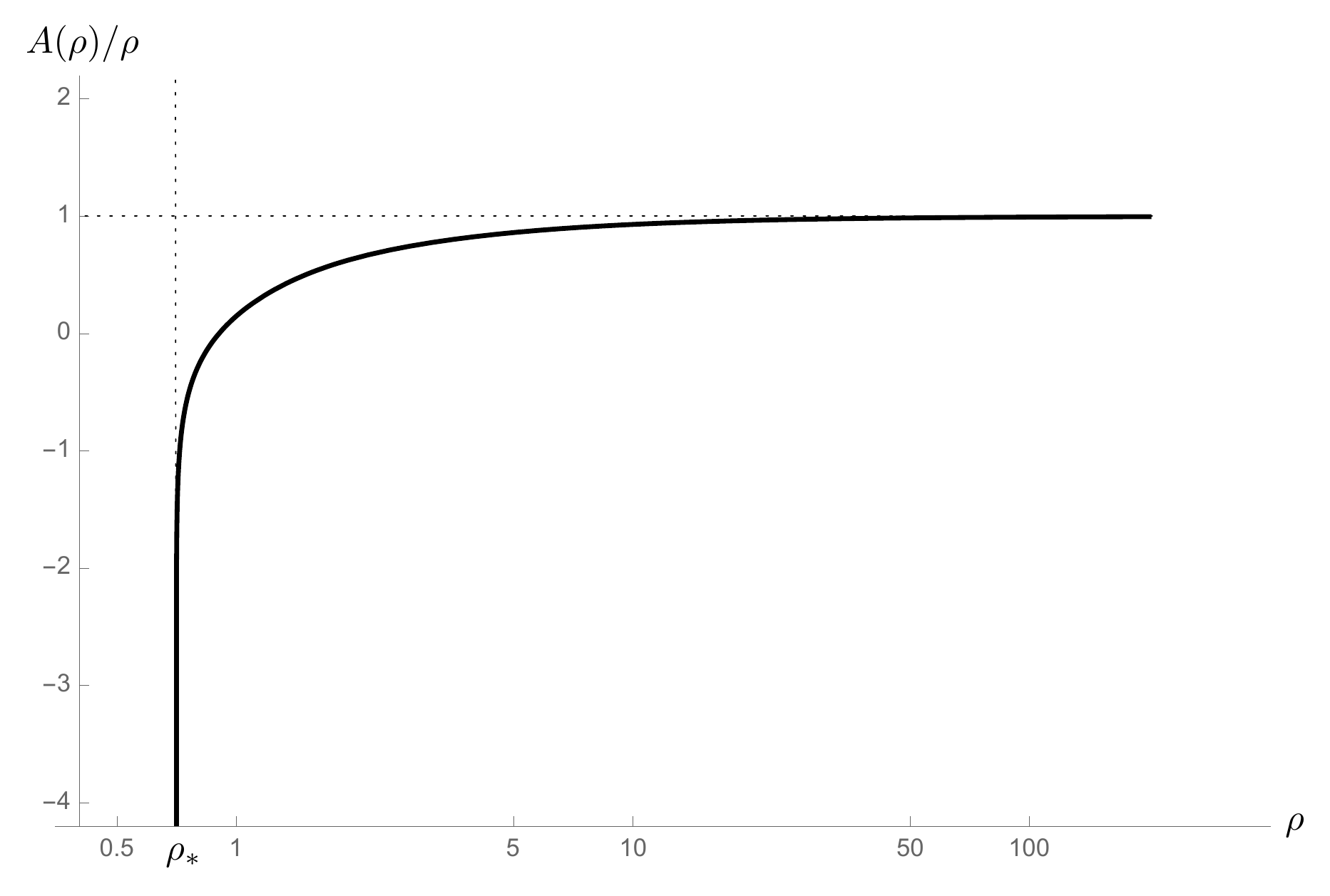}
\caption{A typical solution.  Here $b=1$.  On the horizontal axis is the $\rho$ coordinate on a logarithmic scale, and on the vertical axis is the scale factor $A(\rho)$, normalized by $\rho$ so that it can be seen that $A\to \rho$ as $\rho\to \infty$, corresponding to asymptotically $\mathrm{AdS}$ boundary conditions.  The singularity occurs at the value $\rho_\ast\approx 0.7$, for which the scale factor goes to $-\infty$. 
}
\end{figure}

A typical solution is shown in Figure \ref{solutionplot}.  In the interior, the background develops a curvature singularity at a value finite value $\rho=\rho_\ast$, determined by $b$, for which $A\rightarrow -\infty$. As $b\to 0$, $\rho_\ast \to 0$ in agreement with the probe calculation.   From the definition \eqref{ucoorddef}, we see that the $u$ coordinate tends to $\infty$ as we approach the singularity, so the $u$ coordinate covers the region from the boundary to the singularity as we range over $u=(0,\infty)$.
This singularity is a naked singularity\footnote{It maps under the domain-wall/cosmology correspondence \cite{Skenderis:2006jq} to a big bang in a spatially open universe.}. The gauge-theory interpretation of this bulk singularity is that the theory is flowing from a conformal fixed point in the UV to a gapped theory at low energies \cite{Kraus:1998hv,Gubser:2000nd}. 
 Since the singularity occurs at a fixed value $\rho_\ast$, we expect the gauge theory in flat space to possess a time-dependent IR cut-off. Transforming to the Poincar\'{e} patch we obtain $z_\ast \sim t \sech \rho_\ast$. Since $1/z$ parametrizes the RG scale, we find expect that the effective cut-off goes like
$\Lambda_{\rm IR} \sim \frac{\cosh\rho_\ast}{t}$.  Note that the geometry of this solution is analogous to the Janus solution of \cite{Bak:2003jk}.

To determine the VEVs of the stress tensor and the operator ${\cal O}$ dual to the scalar field, we must first write the metric and scalar in Fefferman-Graham coordinates \cite{Fefferman:2007rka}, which is always possible in asymptotically $\mathrm{AdS}$ spacetime,
\begin{align}
ds^2
		&= {1\over z^2}\left(dz^2 + \left[ g_{\mu\nu}(x) + z^2 g_{(2)\mu\nu}(x) + z^4 (t_{\mu\nu}(x) + g_{(4)\mu\nu}(x)\log z )  + \cdots\right]dx^\mu dx^\nu\right)\ , \label{FGgmet} \\
	{\phi}(x,z)
		& = \varphi(x) + z^2 \varphi_{(2)}(x) + z^4 (\psi(x)+\varphi_{(4)}(x)\log z) + \cdots \label{FGgrsc1}
\end{align}
The expansion for the scalar, which includes the logarithmic term, is the appropriate one for a massless field in the bulk.   The leading term $g_{\mu\nu}(x)$ is the metric on which the boundary field theory lives, and $\varphi(x)$ is the source of the scalar operator in the boundary theory.  In Appendix \ref{hrenormappe}, we calculate the unambiguous parts of the exact one-point functions of the scalar operator and stress tensor, in the presence of a general source and boundary metric.  The calculation shows that the one-point functions of the scalar operator $\langle\mathcal{O}\rangle$ and the stress tensor $\langle T_{\mu\nu}\rangle$, for zero source and flat background, are determined up to numerical constants by the coefficients of $z^4$ in the Fefferman-Graham expansions of the bulk fields, 
\be \langle\mathcal{O}\rangle=4\psi,\ \ \ \langle T_{\mu\nu}\rangle=2t_{\mu\nu}.\label{vevforms}\ee

{We must find the flat sliced Fefferman-Graham expansion of the metric \eqref{e:janus}. Following \cite{Papadimitriou:2004rz}, we first define a new coordinate $y\in (1,\infty)$ through the definition $y^2 = 1 + u + b u^4$. For $b = 0$ the metric is $\mathrm{AdS}_{5}$, which in the $y$ coordinates reads
\begin{equation}
	ds^2 = \frac{dy^2}{(y^2 -1)^2} + \frac{1}{y^2 - 1} \left[\frac{-d\eta^2+d\vec{x}^2}{\eta^2}\right] \ .
\end{equation}
Our goal is to change to flat-sliced Fefferman-Graham coordinates, and in the $b=0$ case the desired coordinates are the Poincar\'{e} coordinates
\begin{equation}
	ds^2 = \frac{dz^2 - dt^2 + d\vec{x}^2}{z^2},
\end{equation}
and the coordinate transformation is (recalling that $(-t) > z$)
\begin{equation}
	y = \frac{(-t)}{\sqrt{t^2 - z^2}}, \quad \eta = \sqrt{t^2 - z^2} \ .
\end{equation}
For $b=c^2/12\neq 0$ we can work perturbatively in $c$.  First solve the equation $y^2 = 1 + u + b u^4$ to linear order in $b$, 
\begin{equation}
	u = y^2 - 1 - b(y^2 - 1)^4 + \mathcal{O}(b^2) \ ,
\end{equation}
and use this to write the metric in terms of the $y$ variable to linear order in $b$,
\begin{equation}
	ds^2 = \frac{dy^2}{(y^2-1)^2}\left[1-6b(y^2 - 1)^{3} + \mathcal{O}(b^2)\right] + \frac{1}{y^2 - 1}\left[1 + b(y^2 -1)^{3} + \mathcal{O}(b^2)\right]ds_{\mathrm{dS}_4}^2 \ .
\end{equation}
Now we need to find, to linear order in $b$, the coordinate transformation which takes us to flat sliced Fefferman-Graham coordinates.   Writing the ansatz, 
\begin{equation}
	y = \frac{t}{\sqrt{t^2 - z^2}} + b f_1(x) + \mathcal{O}(b^2), \quad \eta = \sqrt{t^2 - z^2} + b z f_2(x) + \mathcal{O}(b^2) \ ,
\end{equation}
where $x \equiv(- t)/z$, the functions $f_1(x)$ and $f_2(x)$ can be fixed by demanding that the transformed metric has Fefferman-Graham form. In particular, we demand that the $\mathcal{O}(b)$ terms in $g_{zz}$ and $g_{tz}$ vanish. This gives two conditions which can be solved to give two coupled ordinary differential equations, the solutions of which are

\begin{align}
	f_1(x)
		& = \frac{2+9 x^2-6 x^4-6 x^2 \left(x^2-1\right)^2 \log\left(1-\frac{1}{x^2}\right)}{8 x \left(x^2-1\right)^{7/2}} ,\\
	f_2(x)
		& = \frac{1}{4} \sqrt{x^2-1} \left[\frac{17-42 x^2+24 x^4}{6 \left(x^2-1\right)^3}+4 \log\left(1-\frac{1}{x^2}\right)+\frac{3 \log\left(1-\frac{1}{x^2}\right)}{x^2-1}\right] \ ,
\end{align}
therefore,
\begin{equation}
	ds^2 = \frac{dz^2\left[1 + \mathcal{O}(b^2)\right] + \left[-1 + b \beta_1(x) + \mathcal{O}(b)^2\right] dt^2 + \left[1 + b \beta_2(x) + \mathcal{O}(b)^2\right] d\vec{x}^2}{z^2} \ ,
\end{equation}
where
\begin{align}
	\beta_1(x)
		& = \frac{1}{4} z^2 \left[\frac{-6 t^4+9 t^2 z^2-2 z^4}{t^2\left(t^2- z^2\right)^2}-\frac{6 \log\left(1-\frac{z^2}{t^2}\right)}{z^2}\right], \\
	\beta_2(x)
		& = \frac{6 t^4 z^2-15 t^2 z^4+11 z^6+6 \left(t^2-z^2\right)^3 \log\left(1-\frac{z^2}{t^2}\right)}{12 \left(-t^2+z^2\right)^3}  \ .
\end{align}
We have chosen the integration constants in the solutions so that $\beta_1(x)$ is strictly real and $\beta_2(x)$ does not contain a constant term. 

Thus, the Fefferman-Graham expansion of the metric \eqref{e:janus} to order $c^2$ is
\begin{align}
	ds^2
		& = \frac{dz^2 + {g}_{tt}(z/t)dt^2 + {g}_{11}(z/t)d\vec{x}^2}{z^2} \ ,
\end{align}
where the functions ${g}_{11}(z/t)$ and ${g}_{tt}(z/t)$ are
\begin{align}
	{g}_{tt}(z/t)
		& = -1 + \frac{b}{8}\left(\frac{z}{t}\right)^8 + \mathcal{O}\left(\left(\frac{z}{t}\right)^{10}\right) ,\\
	{g}_{11}(z/t)
		& = 1 - \frac{b}{8}\left(\frac{z}{t}\right)^8 + \mathcal{O}\left(\left(\frac{z}{t}\right)^{10}\right) \ .
\end{align}
Comparing to \eqref{FGgmet}, \eqref{vevforms}, we see that the absence of a $z^4$ coefficient in the Fefferman-Graham expansion of the spatial metric reveals that all components of the one-point function $\langle T_{\mu\nu} \rangle$ vanish, 
\be \la T_{\mu\nu}\ra=0.\ee

Integrating the scalar field equation \eqref{e:scalar} with the help of the chain rule
\begin{equation}
	{\phi} = -\frac{c}{2}\int dx \, \frac{du}{dx} \frac{u}{\sqrt{1+u + b u^4}} \ ,
\end{equation}
and using the asymptotic expansion for $u$ in terms of $x$ we can obtain the $z$-dependence of ${\phi}$,
\begin{equation}
	{\phi} = const. - \frac{c}{4}\left(\frac{z}{t}\right)^4 + \mathcal{O}\left((z/t)^{6}\right).
\end{equation}
One can show by Taylor expanding that the next power in $c$ is $\mathcal{O}(c^3(z/t)^{12})$ which is subdominant, as are all higher powers of $c$. Similarly, the $c^4$ corrections to the metric are  at least $\mathcal{O}((z/t)^8)$, and so cannot contribute to the stress-tensor VEV.}

The pseudo-conformal solutions studied in \cite{Creminelli:2010ba,Hinterbichler:2011qk} have the property that the energy density vanishes but the pressure does not, instead getting a profile $p\sim 1/t^4$ consistent with the symmetries.  Once coupled to gravity, this makes for a very stiff equation of state which is essential for the scalar field component to dominate over other cosmological components such as matter, radiation, curvature, anisotropy, etc.   This serves to empty out and smooth the universe and address the standard puzzles of big bang cosmology without the need for an exponentially expanding spacetime.
Here instead we find a vanishing pressure.  The difference is due to the fact that the AdS/CFT computation is computing the improved stress tensor of the CFT, which is traceless, and which one would use to couple the theory to gravity in a Weyl invariant manner.  The pseudo-conformal solutions, on the other hand, are coupled minimally to gravity, and so the stress tensor which gets a $\sim 1/t^4$ profile is the minimal stress tensor which one uses to couple the theory minimally to gravity.   A CFT on flat space with a VEV $\la {\cal O}\ra\sim 1/t^\Delta$ is equivalent via a Weyl transformation to a CFT on de Sitter with a constant VEV, so it is important that the pseudo-conformal scenario is minimally coupled to gravity rather than conformally coupled, otherwise it would be equivalent to inflation via a Weyl transformation.

\section{Discussion}
The holography of flat domain walls, and more recently their AdS-sliced counterparts, has been the subject of much study over the last decade.  
Here we have studied de Sitter sliced domain walls, arguing that the natural dual theory can be identified with the pseudo-conformal universe. We have calculated scalar and tensor one-point functions in a specific geometry analogous to the Janus solution of \cite{Bak:2003jk}, which realizes the spontaneous symmetry breaking pattern $\Orth(2,4)\rightarrow \Orth(1,4)$.

The domain-wall background in which we calculated correlation functions is singular in the interior of the spacetime. It would be desirable to find a background that is regular throughout. Another pressing problem is to check stability of dS-sliced domain walls. There exist positive-energy arguments which guarantee the classical stability of flat and AdS-sliced domain walls \cite{Freedman:2003ax}, however these arguments rely on the existence of `fake supersymmetry' and do not directly apply to dS-sliced domain walls \cite{Skenderis:2006jq}. 

An issue with the conformal universe scenario is that it requires weight-0 operators in order to generate a scale invariant spectrum.  Naively there is a problem because the unitarity bound $\Delta \geq 1$ in four dimensions \cite{Mack:1975je} prevents the existence of such operators.  One of the assumptions underlying this bound is that there is a conformal vacuum.  This corresponds to the existence of an $\mathrm{AdS}$ vacuum solution in the bulk theory.  It would be interesting to investigate whether it is possible to have bulk theories which are $\mathrm{AdS}$ invariant and yet have no $\mathrm{AdS}$ vacuum, which could be dual to conformal theories with no conformal vacuum and a chance to allow dimension $0$ operators. 

It would be interesting to study higher-point correlation functions.  These satisfy Ward identities for the spontaneously broken symmetries, known in cosmology as consistency relations \cite{Hinterbichler:2013dpa}.  The consistency relations for pseudo-conformal correlators have been studied in \cite{Creminelli:2012qr}, and it should be possible to understand these relations holographically. 

Finally, it would be interesting to consider the holographic dual to turning on 4D gravity in the boundary theory.  In the pseudo-conformal scenario, gravity becomes important at late times.  For example, in the main example of \cite{Hinterbichler:2011qk} where a weight 1 field drives the scenario, the approximation of ignoring gravity is good until $\langle\mathcal{O}\rangle \sim M_{\rm P}$, at which point the effect of higher-dimension, Planck-suppressed operators becomes important.  Dynamical gravity can be incorporated in the boundary by moving the boundary slightly into the bulk, to a cut-off surface at $z = \epsilon$.  Local counter-terms which previously diverged in the $\epsilon \to 0$ limit are now regarded as finite kinetic terms for the graviton \cite{ArkaniHamed:2000ds,PerezVictoria:2001pa}. We leave these problems to future work.

\vspace{.5cm}
\noindent
{\bf Acknowledgments}:  The authors would like to thank Vijay Balasubramanian, Steven Gubser, Monica Guica, Justin Khoury, Paul McFadden, Ioannis Papadimitriou and Zain Saleem for helpful discussions and correspondence.
Research at Perimeter Institute is supported by the Government of Canada through Industry Canada and by the Province of Ontario through the Ministry of Economic Development and Innovation. This work was made possible in part through the support of a grant from the John Templeton Foundation.  The opinions expressed in this publication are those of the authors and do not necessarily reflect the views of the John Templeton Foundation (K.H.).  The work of M.T. was supported in part by the US Department of Energy, and J.S. and M.T. are supported in part by NASA ATP grant NNX11AI95G.

\appendix

\section{Holographic Renormalization and Exact One Point Functions\label{hrenormappe}}

In this Appendix, we calculate the holographic one point functions of the scalar plus gravity system for a scalar with no potential, including all the appropriate holographic renormalizations as discussed in \cite{Henningson:1998gx,de Haro:2000xn,Bianchi:2001kw,Skenderis:2002wp}.  This calculation has not been 
explicitly performed elsewhere, to our knowledge.
 
At the boundary our spacetime approaches AdS, so in a neighborhood of the boundary it is possible to transform to Fefferman-Graham coordinates \cite{Fefferman:2007rka},
\begin{align}
ds^2
		&= {1\over z^2}\left(dz^2 + \tilde{g}_{\mu\nu}(x,z)dx^\mu dx^\nu\right)\ , \ \ \ \ \phi(x,z)=\tilde\varphi(x,z),\nn\\
	\tilde{g}_{\mu\nu}(x,z) 
		& = g_{\mu\nu}(x) + z^2 g_{(2)\mu\nu}(x) + z^4 (t_{\mu\nu}(x) + g_{(4)\mu\nu}(x)\log z )  + \cdots, \label{FGgmetA} \\
	{\tilde\varphi}(x,z)
		& = \varphi(x) + z^2 \varphi_{(2)}(x) + z^4 (\psi(x)+\varphi_{(4)}(x)\log z) + \cdots \, . \label{FGgrsc}
\end{align}
The leading term $g_{\mu\nu}(x)$ is the metric on which the boundary field theory lives (which for our purposes we will eventually take to be $\eta_{\mu\nu}$), and $\varphi(x)$ is the source of the dimension 4 scalar operator ${\cal O}$ in the boundary theory (which for our purposes we will eventually take to vanish).  The sub-leading terms are determined by solving the equations of motion. 

The purpose of the following is to show that the one-point functions of the scalar operator $\langle\mathcal{O}\rangle$ and the stress tensor $\langle T_{\mu\nu}\rangle$ are determined (up to numerical constants) by the coefficients of $z^4$ in the Fefferman-Graham expansions of the bulk fields ($\psi$ and $t_{\mu\nu}$), and to quantify the ambiguities in these one-point functions due to different renormalization schemes.

In AdS/CFT, the generating function of boundary field theory correlators as a function of $g_{\mu\nu}(x),\varphi(x)$ is given by the bulk action evaluated on \eqref{FGgrsc} \cite{Witten:1998qj,Gubser:1998bc}. Breaking the bulk metric into sideways ADM  \cite{Arnowitt:1962hi} coordinates with respect to $z$, with lapse $N$, shift $N^\mu$ and spatial metric $\gamma_{\mu\nu}$, 
\be G_{\rm bulk}= \left(\begin{array}{c|c} N^2+N^\mu N_{\mu}  & N_{\mu} \\ \hline N_{\mu} & \gamma_{\mu\nu}  \end{array}\right),\ee
the bulk action \eqref{e:effectiveaction} including the boundary term becomes
\be S={1\over 2} \int_0^\infty dz \int  d^4x  \sqrt{-\gamma}N\left(R[\gamma]+K^2-K_{\mu\nu}^2+12-\left({\cal L}_n\phi\right)^2-(\nabla\phi)^2\right)\label{ADMaction} \ ,
\ee
where the extrinsic curvature and Lie derivative in the normal direction is
\be 
K_{\mu\nu}={1\over 2}{\cal L}_n g_{\mu\nu}={1\over 2N}\left( \gamma'_{\mu\nu}-\nabla_\mu N_\nu-\nabla_\nu N_\mu\right),\ \  {\cal L}_n\phi={1\over N}\left(\phi'-N^\mu\partial_\mu\phi\right) \ ,
\ee
 primes denote derivatives with respect to $z$, and index movements and covariant derivatives are with respect to the spatial metric $\gamma_{\mu\nu}$.

In the Fefferman-Graham coordinates \eqref{FGgrsc}, we have
\be 
N={1\over z},\ \ \ \ N^\mu=0,\ \ \ \ \gamma_{\mu\nu}={1\over z^2} \tilde g_{\mu\nu} \ .
\label{FGgaugeADM}
\ee

To obtain the equations of motion in Fefferman-Graham gauge we first vary with respect to $N$, $N^\mu$, $\gamma_{\mu\nu}$ and $\phi$, then impose \eqref{FGgaugeADM}, yielding
\bea  
&&\tilde g^{\rho\sigma}\tilde \nabla_\rho \tilde g'_{\mu\sigma}-\tilde \nabla_\mu  \Tr (\tilde g^{-1}\tilde g')=2\phi' \partial_\mu\phi,\label{NmuequationS} \\
&& \Tr(\tilde g^{-1}\tilde g'')-{1\over 2} \Tr(\tilde g^{-1}\tilde g'\tilde g^{-1}\tilde g')-{1\over z}\Tr(\tilde g^{-1}\tilde g')=-2\phi'^2,\label{NequationS} \\
&& \tilde g''_{\mu\nu}-(\tilde g'\tilde g^{-1}\tilde g')_{\mu\nu}+{1\over 2}\Tr(\tilde g^{-1}\tilde g') \tilde g'_{\mu\nu}-2R_{\mu\nu}(\tilde g)-{1\over z}\Tr(\tilde g^{-1}\tilde g') \tilde g_{\mu\nu}-{3\over z}\tilde g'_{\mu\nu}= -2\partial_\mu\phi\partial_\nu\phi,\nn \\ \label{gammaequationS}  \\ 
&& \phi''+{1\over 2} \Tr(\tilde g^{-1}\tilde g')\phi'-{3\over z}\phi'+\tilde\square\phi=0 \ .
\label{phieql}
\eea
Here \eqref{NmuequationS} is the $N^\mu$ equation,  \eqref{NequationS} is a linear combination of the $N$ equation and the trace of the $\gamma_{\mu\nu}$ equation (with the coefficient chosen so as to cancel the scalar curvature term $ R[\tilde g]$), \eqref{gammaequationS} is a linear combination of the $\gamma_{\mu\nu}$ equation, $\gamma_{\mu\nu}$ times the $N$ equation, and $\gamma_{\mu\nu}$ times the trace of the $\gamma_{\mu\nu}$ equation (with the coefficients chosen to cancel the $\Tr(\tilde g'')$ term and the $R[\tilde g] \tilde g_{\mu\nu}$ term), and \eqref{phieql} is the $\phi$ equation.

We solve these equations of motion order by order in $z$ by inserting the Fefferman-Graham expansion \eqref{FGgrsc} into \eqref{NmuequationS}, \eqref{NequationS}, \eqref{gammaequationS}, \eqref{phieql} and collecting like powers of $z$.
The important parts of the expansion are the following:
the lowest two orders in the expansion of \eqref{NmuequationS},
\bea && \nabla^\nu g^{(2)}_{\mu\nu}-\nabla_\mu g^{(2)}-2\varphi^{(2)}\nabla_\mu \varphi =0 \ ,\label{NmuequationS1}\\
&&  \nabla^\nu t_{\mu\nu}-\nabla_\mu t +{3\over 4}g^{(2)\rho\sigma}\nabla_\mu g^{(2)}_{\rho\sigma}+{1\over 4}g^{(2)\rho}_{\ \mu}\nabla_\rho g^{(2)}-{1\over 2}g^{(2)\rho\sigma}\nabla_\rho g^{(2)}_{\mu\sigma}-{1\over 2}g^{(2)}_{\mu\rho}\nabla_\sigma g^{(2)\rho\sigma}\nn\\ 
&& +{1\over 4}\left(\nabla^\nu g^{(4)}_{\mu\nu}-\nabla_\mu g^{(4)}\right)-{1\over 2}\varphi^{(4)}\nabla_\mu\varphi-2\psi\nabla_\mu\varphi+\log z\left(\nabla^\nu g^{(4)}_{\mu\nu}-\nabla_\mu g^{(4)}-2\varphi^{(4)}\nabla_\mu\varphi\right)=0 \ , \nn 
\label{NmuequationS2}
\eea
the lowest order in the expansion of \eqref{NequationS},
\bea && t-{1\over 4}g^{(2)}_{\mu\nu}g^{(2)\mu\nu}+{3\over 4} g^{(4)}+(\varphi^{(2)})^2+g^{(4)}\log z =0 \ , 
\label{NequationS1}
\eea
the lowest two orders in the expansion of \eqref{gammaequationS},
\bea && 2 g^{(2)}_{\mu\nu}+ g^{(2)} g_{\mu\nu}+R_{\mu\nu}-\nabla_\mu\varphi\nabla_\nu\varphi =0 \ ,\label{gammaequationS1}\\
&& g^{(4)}_{\mu\nu}+{1\over 2}g_{\mu\nu}g^{(2)}_{\rho \sigma}g^{(2)\rho\sigma}-g^{(2)}_{\mu \rho}g^{(2)\rho}_{\nu}-{1\over2} \nabla_{(\mu}\nabla^\rho g^{(2)}_{\nu)\rho}+{1\over 4}\nabla^2 g^{(2)}_{\mu\nu}+{1\over 4}\nabla_\mu\nabla_\nu g^{(2)}-{1\over 2}g^{(2)}_{\rho(\mu}R^\rho_{\ \nu)} \nn\\
&&+{1\over 2}g^{(2)\rho \sigma}R_{\mu\rho\nu\sigma}-g_{\mu\nu} t-{1\over 4} g_{\mu\nu} g^{(4)}+\nabla_{(\mu}\varphi\nabla_{\nu)}\varphi^{(2)}-  g_{\mu\nu} g^{(4)}\log z=0 \ ,\label{gammaequationS2}
\eea
and the lowest two orders in the expansion of \eqref{phieql},
\bea 
&& \varphi^{(2)}-{1\over 4}\square\varphi=0 \ ,\label{phieql1}\\
&& \varphi^{(4)}+{1\over 2}g^{(2)}\varphi^{(2)}+{1\over 4}\square\varphi^{(2)}+{1\over 8}\nabla_\mu g^{(2)}\nabla^\mu\varphi-{1\over 4}\nabla_\mu g^{(2)\mu\nu}\nabla_\nu\varphi -{1\over 4}g^{(2)\mu\nu}\nabla_\mu\varphi\nabla_\nu\varphi=0 \ .\nn 
\label{phieql2}
\eea
In these expressions, all index raising/lowering, covariant derivatives, curvatures, etc. are with respect to $g_{\mu\nu}$.

Equation \eqref{gammaequationS1} determines $g^{(2)}_{\mu\nu}$; by taking a trace and reinserting the result in to the equation we find
\be 
g^{(2)}_{\mu\nu}=-{1\over 2}\left(R_{\mu\nu}-{1\over 6}Rg_{\mu\nu}-\nabla_\mu\varphi\nabla_\nu\varphi+{1\over 6}g_{\mu\nu}(\nabla\varphi)^2\right) \ .
\label{g2solS}
\ee
Equation \eqref{phieql1} determines ${\varphi^{(2)}}$,
\be 
\varphi^{(2)}={1\over 4}\square\varphi\label{phi2sol} \ ,
\ee
and \eqref{NmuequationS1} is automatically satisfied by \eqref{g2solS}, \eqref{phi2sol}, as can be checked using the Bianchi identity.

Equation \eqref{phieql2} determines $\varphi^{(4)}$,
\bea 
\varphi^{(4)}&&={1\over 2}g^{(2)}\varphi^{(2)}-{1\over 4}\square\varphi^{(2)}-{1\over 8}\nabla_\mu g^{(2)}\nabla^\mu\varphi+{1\over 4}\nabla_\mu g^{(2)\mu\nu}\nabla_\nu\varphi +{1\over 4}g^{(2)\mu\nu}\nabla_\mu\varphi\nabla_\nu\varphi\nn\\
&&=-{1\over 16}\square^2\varphi+{1\over 6}\nabla^\mu\varphi\nabla^\nu\varphi\nabla_\mu\nabla_\nu\varphi+{1\over 12}(\nabla\varphi)^2\square\varphi-{1\over 8}R^{\mu\nu}\nabla_\mu\varphi\nabla_\nu\varphi+{1\over 24}R(\nabla\varphi)^2-{1\over 48}\nabla^\mu R\nabla_\mu\varphi \ .\nn
\eea

Moving to \eqref{NequationS1}, the logarithm term must vanish separately, which tells us that $g^{(4)}_{\mu\nu}$ is traceless,
\be 
g^{(4)}=0\label{g4tracelessS} \ ,
\ee
and the remainder of the equation then determines the trace of $t_{\mu\nu}$,
\bea 
t={1\over 4}g^{(2)}_{\mu\nu}g^{(2)\mu\nu}-(\varphi^{(2)})^2={1\over 16}R_{\mu\nu}^2-{1\over 72}R^2+{1\over 36}R(\nabla\varphi)^2-{1\over 8}R^{\mu\nu}\nabla_\mu\varphi\nabla_\nu\varphi-{1\over 16}(\square\varphi)^2+{7\over 144}(\nabla\varphi)^4\label{ttraceS} \ .\nn 
\eea

The logarithm term of \eqref{NmuequationS2} must also vanish separately, which tells us the divergence of $g^{(4)}_{\mu\nu}$,
\be 
\nabla^\nu g^{(4)}_{\mu\nu}=2\varphi^{(4)}\nabla_\mu\varphi \label{g4divS} \ ,
\ee
and the rest of \eqref{NmuequationS2} then determines the divergence of $t_{\mu\nu}$,
\bea 
\nabla^\nu t_{\mu\nu}=2\psi\nabla_\mu\varphi-\varphi^{(2)}\nabla_\mu\varphi^{(2)}-{1\over 4}g^{(2)\rho\sigma}\nabla_\mu g^{(2)}_{\rho\sigma}-{1\over 4}g^{(2)\rho}_{\ \mu}\nabla_\rho g^{(2)}+{1\over 2}g^{(2)\rho\sigma}\nabla_\rho g^{(2)}_{\mu\sigma}+{1\over 2}g^{(2)}_{\mu\rho}\nabla_\sigma g^{(2)\rho\sigma} \ .\nn
\eea

Now, equation \eqref{gammaequationS2} determines $g^{(4)}_{\mu\nu}$,
 \bea 
 g^{(4)}_{\mu\nu}= && -{1\over 4}g_{\mu\nu}g^{(2)}_{\rho \sigma}g^{(2)\rho\sigma}+g^{(2)}_{\mu \rho}g^{(2)\rho}_{\nu}+{1\over2} \nabla_{(\mu}\nabla^\rho g^{(2)}_{\nu)\rho}-{1\over 4}\nabla^2 g^{(2)}_{\mu\nu}-{1\over 4}\nabla_\mu\nabla_\nu g^{(2)}+{1\over 2}g^{(2)}_{\rho(\mu}R^\rho_{\ \nu)} \nn\\
&&-{1\over 2}g^{(2)\rho \sigma}R_{\mu\rho\nu\sigma}-g_{\mu\nu} (\varphi^{(2)})^2-\nabla_{(\mu}\varphi\nabla_{\nu)}\varphi^{(2)} \ ,\nn
\eea 
which is consistent with \eqref{g4tracelessS} and \eqref{g4divS}.

Note that $g^{(4)}_{\mu\nu}$ and $\varphi^{(4)}$ are functional derivatives 
\be
 g^{(4)}_{\mu\nu}=-{2\over \sqrt{-g}}{\delta S_W\over \delta g^{\mu\nu}},\ \ \ \ \  \ 2\varphi^{(4)}={1\over \sqrt{-g}}{\delta S_W\over \delta\varphi} \ ,
\ee
of an action
\bea 
&&S_W={1\over 8}\int d^4x\,\sqrt{-g}\left[-{1\over 4}C_{\mu\nu\rho\sigma}C^{\mu\nu\rho\sigma}-{1\over 2}(\square\varphi)^2-{1\over 3}(\nabla\varphi)^4+R^{\mu\nu}\nabla_\mu\varphi\nabla_\nu\varphi-{1\over 3}R(\nabla\varphi)^2\right] \ , \nn \\
\label{Weyllag}
\eea
with $C_{\mu\nu\rho\sigma}$ the Weyl tensor.
The expression \eqref{g4divS} is nothing but the Ward identity for diffeomorphism invariance of this action.  This action is also Weyl invariant, with $\varphi$ transforming with Weyl weight $0$, and  \eqref{g4tracelessS} is the corresponding Ward identity.

We can now evaluate the on shell action by plugging the solution into \eqref{ADMaction}.  The action found in this way is divergent due to the infinite volume of $\mathrm{AdS}_{5}$, and so we define a regulated on-shell action by placing the boundary at $z=\epsilon$,
\bea 
S_\epsilon[g,\varphi]&=& S_{\epsilon,{\rm bulk}}[g,\varphi]+S_{\epsilon,{\rm boundary}}[g,\varphi] \ ,
\eea
where
\bea
S_{\epsilon,{\rm bulk}}[g,\varphi]&=& -4 \int  d^4x \int_\epsilon^\infty dz \, N\sqrt{-\gamma},\\
S_{\epsilon,{\rm boundary}}[g,\varphi]&=&-  \int  d^4x \left[ \sqrt{-\gamma}K\right]_{z=\epsilon} \ .
\label{onshellact} 
\eea
We have simplified the bulk part by using the trace of the bulk Einstein equations, $R[G]-(\partial_A\phi)^2=-20$.
 Expanding for small $\epsilon$, the resulting expression contains local divergent terms and non-local finite terms,
\be  
S_\epsilon[g,\varphi]= \int  d^4x \sqrt{-g}\left[ {1\over \epsilon^{4}}a_4+{1\over \epsilon^{2}}a_{2}+\log \epsilon \, a_0+{\rm finite}+\cdots\right] \ ,
\label{a0gend}
\ee
where the $a$'s are the following local quantities constructed from $g_{\mu\nu}$ and $\varphi$, 
\bea a_4&=& 3, \\
a_2 &=&0 ,\\ 
a_0&=& - g^{(2)}_{\mu\nu}g^{(2)\mu\nu}+{1\over 2} \left(g^{(2)}\right)^2+2t\nn\\ 
&&=-{1\over 8}\left(R_{\mu\nu}R^{\mu\nu}-{1\over 3}R^2\right)-{1\over 8}(\square\varphi)^2-{1\over 12}(\nabla\varphi)^4+{1\over 4}R^{\mu\nu}\nabla_\mu\varphi\nabla_\nu\varphi-{1\over 12}R(\nabla\varphi)^2.\nn\\
\eea
Note that $a_0$ is proportional, up to a total derivative, to the Lagrangian of \eqref{Weyllag}.

Counterterms must be chosen to cancel the infinite terms, and are ambiguous up to local finite terms,
\be  S_{c.t.}[g,\varphi]= \int  d^4x \sqrt{-g}\left[ -{1\over \epsilon^{4}}a_4-{1\over \epsilon^{2}}a_{2}-\log \epsilon \, a_0+{\rm local\ finite}\right].\label{countertermsgen} \ \ee
The generating function is the regulated generating functional minus the counter terms and is finite,
\be 
S[g,\varphi]=\lim_{\epsilon\rightarrow 0}\left(S_\epsilon[g,\varphi]+S_{c.t.}[g,\varphi]\right) \ .
\ee
The stress tensor and scalar VEVs can then be calculated by functionally differentiating, and the result is ambiguous up to functional derivatives of local finite terms.

Calculating directly in this way would require finding the complicated non-local finite part of $S[g,\varphi]$.  
Instead we can proceed indirectly.  Start by defining $\la T_{\mu\nu}\ra^\epsilon$ as the stress tensor obtained by functionally differentiating the regulated action, before subtracting counterterms.  It is a function of the full boundary metric $\tilde g_{\mu\nu}(x,\epsilon)$ and boundary scalar $\tilde \varphi(x,\epsilon)$.
Since the bulk variation is always just the equation of motion, in varying \eqref{ADMaction} the only contribution on shell is a boundary term at the cutoff $z=\epsilon$,
\be \delta S_\epsilon=-{1\over 2} \int d^4x\sqrt{-\gamma}\left(K_{\mu\nu}-K\gamma_{\mu\nu}\right)\delta\gamma^{\mu\nu}\bigg|_{z=\epsilon}.\ee
Thus we have,
\bea 
\la T_{\mu\nu}\ra^\epsilon&=&-{2\over \sqrt{-\tilde g}}{\delta S_\epsilon[\tilde g,\tilde\varphi]\over \delta \tilde g^{\mu\nu}}=-{1\over \epsilon^{2}}{2\over \sqrt{-\gamma}}{\delta S_\epsilon[\gamma,\phi]\over \delta \gamma^{\mu\nu}}={1\over \epsilon^{2}}\left(K_{\mu\nu}-K\gamma_{\mu\nu}\right)\bigg|_{\epsilon}\nn\\
&=&{1\over 2\epsilon^{3}}\left(\tilde g'_{\mu\nu}-\tilde g_{\mu\nu}\Tr(\tilde g^{-1}\tilde g')+{6\over z}\tilde g_{\mu\nu}\right)\bigg|_{\epsilon} \ .
\label{regulatedtmnS}
\eea
The variation of the regulated action \eqref{ADMaction} with respect to the scalar gives the regulated scalar one-point function,
\bea 
\delta S&&= -{1\over 2}\int  d^4x\ \sqrt{-\gamma}{1\over N}\left(\phi'-N^\mu\partial_\mu\phi\right)\delta\phi'=\int d^4x\ {1\over \epsilon^{3}}\sqrt{-\tilde g}\,\phi'\delta\phi =\int d^4x\ {1\over \epsilon^{3}}\sqrt{-\tilde g}\,\tilde\varphi'\delta\tilde\varphi , \nn \\
\label{varrega}
\eea
\be 
\la {\cal O}\ra^\epsilon={1\over \sqrt{-\tilde g}}{\delta S\over\delta \tilde\varphi}={1\over \epsilon^{3}}\tilde\varphi' \ .\label{regvev}
\ee
Evaluating these to finite order using the Fefferman-Graham expansion solutions, we have
\bea  
\la T_{\mu\nu}\ra^\epsilon&=& {3\over \epsilon^4}g_{\mu\nu}+{1\over \epsilon^2}\left(4g^{(2)}_{\mu\nu}- g_{\mu\nu}g^{(2)}\right)+{5}\log \epsilon\,g^{(4)}_{\mu\nu}+{5}t_{\mu\nu}+{1\over 2}g^{(4)}_{\mu\nu}\nn\\
&&-g^{(2)}_{\mu\nu}g^{(2)}+g_{\mu\nu}g^{(2)}_{\rho\sigma}g^{(2)\rho\sigma}-2g_{\mu\nu}t\ ,
\label{tregex} \\
\la {\cal O}\ra^\epsilon 
&=&{2\over \epsilon^2}\varphi^{(2)}+4\varphi^{(4)}\log z+4\psi+\varphi^{(4)} \ .
\eea

Next we define the counter-term stress tensor and scalar one-point function, which are obtained by differentiating the counterterm action with respect to $\tilde g^{\mu\nu}$,
\be  
T_{\mu\nu}^{c.t.}=-{2\over \sqrt{-\tilde g}}{\delta S_{c.t.}[\tilde g,\tilde\varphi]\over \delta \tilde g^{\mu\nu}} \ ,
\ee
\be \la {\cal O}\ra^\epsilon_{c.t}={1\over \sqrt{-\tilde g}}{\delta S[\tilde g,\tilde \varphi]_{c.t.}\over\delta \tilde\varphi}.
\ee
To calculate these, we must express the counterterm action in terms of $\tilde g_{\mu\nu},\tilde\varphi$ rather than $g_{\mu\nu},\varphi$ by inverting the Fefferman-Graham expansion up to the required order.  We write
\bea g_{\mu\nu}&=&\tilde g_{\mu\nu}+z^2\tilde g^{(2)}_{\mu\nu}+z^4\tilde g^{(4)}_{\mu\nu}+\cdots, \nn\\
\varphi &=&\tilde \varphi+z^2\tilde \varphi^{(2)}+z^4\tilde \varphi^{(4)}+\cdots \label{invertansatzS},
\eea
then plug in the Fefferman-Graham expansion and equate powers of $z$ to obtain
\bea \tilde g^{(2)}_{\mu\nu}&=&-g^{(2)}_{\mu\nu}\bigg|_{\tilde g,\tilde\varphi},\nn\\
 \tilde g^{(4)}_{\mu\nu}&=&-{\delta  g^{(2)}_{\mu\nu}\over \delta g_{\rho\sigma}}g^{(2)}_{\rho\sigma}-{\delta  g^{(2)}_{\mu\nu}\over \delta \varphi}\varphi^{(2)}-\left(t_{\mu\nu}+g^{(4)}_{\mu\nu}\log z\right)\bigg|_{\tilde g,\tilde\varphi},\nn\\
 \tilde \varphi^{(2)}&=&-\varphi^{(2)}\bigg|_{\tilde g,\tilde\varphi},\nn\\
 \tilde \varphi^{(4)}&=&-{\delta  \varphi^{(2)}\over \delta g_{\mu\nu}}g^{(2)}_{\mu\nu}-{\delta  \varphi^{(2)}\over \delta \varphi}\varphi^{(2)}-\left(\psi+\varphi^{(4)}\log z\right)\bigg|_{\tilde g,\tilde\varphi}.
\eea

 Expressing the counterterm action \eqref{countertermsgen} in terms of $\tilde g_{\mu\nu}$ and $\tilde \varphi$, we find,
\be 
S_{c.t.}[\tilde g,\tilde\varphi]= \int  d^4x \sqrt{-\tilde g}\left[-{3\over \epsilon^4}+{1\over 4 \epsilon^2}\left(-\tilde R+(\tilde\nabla\tilde \varphi)^2\right)-\log\epsilon\,a_0[\tilde g,\tilde\varphi]+{\rm local\ finite}\right] \ .
\ee
Varying this, we obtain the counterterm stress tensor and VEV in terms of $\tilde g_{\mu\nu}$ and $\tilde \varphi$,
\bea 
\la T_{\mu\nu}\ra^{c.t.}[\tilde g,\tilde\varphi]&=&-{2\over \sqrt{-\tilde g}}{\delta S_{c.t.}[\tilde g,\tilde\varphi]\over \delta \tilde g^{\mu\nu}} \nn\\
&=&-{3\over \epsilon^4}\tilde g_{\mu\nu}+{1\over 2\epsilon^2}\left(\tilde R_{\mu\nu}-{1\over 2}\tilde R\tilde g_{\mu\nu}-\tilde\nabla_\mu\tilde\varphi\tilde\nabla_\nu\tilde\varphi+{1\over 2}\tilde g_{\mu\nu}(\tilde\nabla\tilde\varphi)^2\right)-2\log\epsilon g^{(4)}(\tilde g,\tilde \varphi)\nn\\ 
&&+\delta({\rm local\ finite}) \ ,\nn\\
\la  {\cal O}\ra^\epsilon_{c.t}[\tilde g,\tilde\varphi]&=&{1\over \sqrt{-\tilde g}}{\delta S_{c.t.}[\tilde g,\tilde \varphi]\over\delta \tilde\varphi} \nn\\
&=&-{1\over 2\epsilon^2}\tilde \square\tilde\varphi-4\log \epsilon \varphi^{(4)}(\tilde g,\tilde\varphi)+\delta({\rm local\ finite}) \ .
\eea

We now insert the Fefferman-Graham expansion into these in order to express them in terms of $ g_{\mu\nu}$ and $ \varphi$,
\bea 
\la T_{\mu\nu}^{c.t.}\ra[ g,\varphi]&=&-{3\over \epsilon^4}g_{\mu\nu}+{1\over \epsilon^2}\left(2R_{\mu\nu}-{1\over 2}Rg_{\mu\nu}-2\nabla_\mu\varphi\nabla_\nu\varphi+g_{\mu\nu}(\nabla\varphi)^2\right)-5\log\epsilon g^{(4)}_{\mu\nu}\nn\\
&&-{3 } t_{\mu\nu}+g^{(4)}_{\mu\nu} -{1\over 16}g_{\mu\nu}R_{\rho\sigma}R^{\rho\sigma}-{1\over 4}R_{\mu\rho}R_\nu^{\ \rho}+{5\over 24}R_{\mu\nu}R-{1\over 48}g_{\mu\nu}R^2 \nn\\
&& +{1\over 2}R_{(\mu}^{\ \rho}\nabla_{\nu)}\varphi\nabla_\rho\varphi-{5\over 24}R_{\mu\nu}(\nabla\varphi)^2+{1\over 24}g_{\mu\nu}R(\nabla\varphi)^2-{5\over 24}R\nabla_\mu\varphi\nabla_\nu\varphi \nn\\
&&-{1\over 24}\nabla_\mu\varphi\nabla_\nu\varphi(\nabla\varphi)^2-{1\over 16}g_{\mu\nu}(\square\varphi)^2+{1\over 8}g_{\mu\nu}R^{\rho\sigma}\nabla_\rho\varphi\nabla_\sigma\varphi-{1\over 12}g_{\mu\nu}(\nabla\varphi)^4\nn\\
&&+\delta({\rm local\ finite}) \ ,\nn\\
\la  {\cal O}\ra^\epsilon_{c.t}[ g,\varphi]&=&-{1\over 2\epsilon^2} \square\varphi -4\log\epsilon\varphi^{(4)}-{1\over 24}R\square \varphi+{1\over 24}(\nabla\varphi)^2\square\varphi+2\varphi^{(4)} +\delta({\rm local\ finite}) \ .\nn
\eea

The renormalized stress tensor and one point function are now given by
\be 
\la T_{\mu\nu}\ra=\lim_{\epsilon\rightarrow 0}\left(T_{\mu\nu}^\epsilon[ g,\varphi]+T_{\mu\nu}^{c.t.}[ g,\varphi]\right) \ ,
\label{finitestS}
\ee

\be 
\la {\cal O}\ra=\lim_{\epsilon\rightarrow 0}\left(\la {\cal O}\ra^\epsilon+\la {\cal O}\ra^\epsilon_{c.t}[g,\varphi]\right) \ ,
\label{finiteVEVs}
\ee
and should be finite.  Because of the ambiguity of local finite contributions to the counterterms \eqref{countertermsgen}, the stress tensor is always ambiguous up to terms which are functional derivatives of local terms.
Evaluating \eqref{finitestS} and \eqref{finiteVEVs}, the divergent pieces all cancel, as they should, yielding
\bea 
\la T_{\mu\nu}\ra&=&2 t_{\mu\nu}+{3\over 2}g^{(4)}_{\mu\nu}+{1\over 16}g_{\mu\nu}R_{\rho\sigma}R^{\rho\sigma}-{1\over 4}R_{\mu\rho}R_\nu^{\ \rho}+{1\over 8}R_{\mu\nu}R-{5\over 144}g_{\mu\nu}R^2 \nn\\
&& +{1\over 2}R_{(\mu}^{\ \rho}\nabla_{\nu)}\varphi\nabla_\rho\varphi-{1\over 8}R_{\mu\nu}(\nabla\varphi)^2+{5\over 72}g_{\mu\nu}R(\nabla\varphi)^2-{1\over 8}R\nabla_\mu\varphi\nabla_\nu\varphi \nn\\
&&-{1\over 8}\nabla_\mu\varphi\nabla_\nu\varphi(\nabla\varphi)^2+{1\over 16}g_{\mu\nu}(\square\varphi)^2-{1\over 8}g_{\mu\nu}R^{\rho\sigma}\nabla_\rho\varphi\nabla_\sigma\varphi+{1\over 36}g_{\mu\nu}(\nabla\varphi)^4\nn\\
&&\delta({\rm local\ finite}) \ ,\nn\\
\la  {\cal O}\ra&=&4\psi+3\varphi^{(4)}-{1\over 24}R\square \varphi+{1\over 24}(\nabla\varphi)^2\square\varphi+3\varphi^{(4)} +\delta({\rm local\ finite}) \ .\nn
\eea
Note that the terms ${3\over 2}g^{(4)}_{\mu\nu}$, $3\varphi^{(4)}$ can be absorbed into $\delta({\rm local\ finite})$ because they stem from the variation of the local action \eqref{Weyllag}. The renormalized VEVs satisfy the Ward identity for diffeomorphisms,
\be 
\nabla^\nu \la T_{\mu\nu}\ra-\la  {\cal O}\ra\nabla_\mu\varphi=0 \ .
\ee

Taking the trace of the stress tensor, we find the Weyl anomaly,
\bea  
{\cal A}&=& g^{\mu\nu}\la T^{}_{\mu\nu}\ra=-a_0 +\Tr\, \delta (\rm local\ finite) \nn\\
&=& -{1\over 16}E_{(4)}+{1\over 16}C_{\mu\nu\rho\sigma}^2+{1\over 8}(\square\varphi)^2+{1\over 12}(\nabla\varphi)^4-{1\over 4}R^{\mu\nu}\nabla_\mu\varphi\nabla_\nu\varphi+{1\over 12}R(\nabla\varphi)^2+ \Tr\, \delta (\rm local\ finite) \ ,\nn
\eea
where $E_{(4)}\equiv R_{\mu\nu\rho\sigma}^2-4R_{\mu\nu}^2+R^2$ is the four dimensional Euler density.

In the case of a flat background with no source (the case of interest in this paper), $g_{\mu\nu}=\eta_{\mu\nu}$, $\varphi=0$, all the curvature and $
\varphi$ terms vanish, and the only ambiguity is a term in the stress tensor proportional to $\eta_{\mu\nu}$, coming from a cosmological constant counterterm, so we have
\be    
\la T_{\mu\nu}\ra=2t_{\mu\nu}+(const.)\eta_{\mu\nu},\ \ \ \ \la  {\cal O}\ra=4\psi~~~~~~~{\rm (flat\ space,\ no\ source)} \ .
\ee
The contribution of the conformal anomaly to the stress-tensor one-point function vanishes on flat space. 
The cosmological constant counterterm can be chosen so that the contribution proportional to $\eta_{\mu\nu}$ vanishes.  Then we see that the one-point functions are indeed given by the order $z^4$ parts of the Fefferman-Graham expansion.

\section{Supergravity Embedding\label{sugraappen}}

The choice of vanishing potential used in Section \ref{backreacsect} can be uplifted to 10 dimensions as a particular truncation of ten-dimensional type IIB supergravity\footnote{
This simple choice of potential has been exploited to construct the AdS-sliced domain wall known as the Janus solution \cite{Bak:2003jk}, originally formulated in ten-dimensional type IIB supergravity.}.

The ten-dimensional Einstein frame equations of motion for the metric, the dilaton and the self-dual five-form in IIB supergravity with all other fields vanishing are
\begin{align}
	 R_{\mu\nu}-\frac{1}{2}g_{\mu\nu}R
 		& =\partial_\mu\phi\partial_\nu\phi -\frac{1}{2}g_{\mu\nu}(\partial\phi)^2 + \frac{1}{24}F_{\mu\rho\lambda\sigma\tau}F_{\nu}^{\phantom{\nu}\rho\lambda\sigma\tau}-\frac{1}{240}g_{\mu\nu}F^2, \\
	\square\phi
		& = 0, \\
	dF_5
		& = 0, \  \\
		\ast F_5
		& = F_5 .
\end{align}
Consider a compactification of the form
\begin{align}
ds^2
	& = d\rho^2 + e^{2A(\rho)}ds^2_{dS_{4}}+
      ds_{S^5}^2, \\
ds^2_{dS_{4}}
	 &=  \frac{-d\eta^2 + d\vec{x}^2}{\eta^2},  \\   
\phi
	& = \phi(\rho), \\
F_5
	&= 4 (\ast\omega_{S^5}
+  \omega_{S^5}) \ ,
\end{align}
where $ \omega_{S^5}$ is the volume form on the internal 5-sphere.  The five-form is manifestly self-dual and satisfies $dF_5 = 0$. 
The scalar equation is solved by 
\begin{equation}\label{scalareqA}
	\phi'(\rho) = c e^{-4A(\rho)} \ ,
\end{equation}
which is the same as the 5D equation \eqref{e:scalar}.
The five form flux acts as a source in the Einstein equation;
we have $R_{\rho\rho}-\frac{1}{2}g_{\rho\rho}R = -10-6e^{-2A(\rho)} + 6A'(\rho)^2$, $F_{\rho ....}F_{\rho}^{\phantom{\rho}....} = -64(4!)$, and the self-duality constraint implies $F^2=0$, so the $\rho$-$\rho$ component of the Einstein equation becomes, upon using \eqref{scalareqA},
\begin{equation}
A'(\rho)^2 = 1+e^{-2A(\rho)}+be^{-8A(\rho)} \ ,
\end{equation}
where $b = c^2/12$.  This is the same as the 5D equation \eqref{e:constraintnew}. In the well-studied $\mathrm{AdS}_5 \times S^5$ compactification of IIB the boundary value of the dilaton is proportional to the Yang-Mills coupling parameter $e^{2\phi} \sim g_{\rm YM}^2$, whose inverse multiplies the $\mathcal{N} = 4$ SYM Lagrangian density.
It is therefore natural to associate a spacetime varying dilaton with a running gauge coupling parameter. 

\section{Comparison with Holographic Defect CFTs}

The pseudo-conformal solutions are essentially Wick rotated holographic interface CFTs, i.e. CFTs in which there is an interface that occurs around the spacelike surface $t=0$.
Consider a 4D CFT with a planar spatial boundary in Euclidean signature. Then the boundary breaks the conformal symmetry $\Orth(1,5)$ to the subgroup $\Orth(1,4)$ of conformal transformations which leave the boundary invariant. The unbroken symmetry group coincides with the isometry group of $\mathrm{dS}_4$, so we expect correlation functions in this theory to be related in a trivial way to those of the pseudo-conformal universe. The one-point functions of scalar operators inserted away from the boundary are fixed by the residual $\Orth(1,4)$ invariance
\begin{equation}
	\langle \mathcal{O}_4(\vec{x},y)\rangle \sim  \frac{1}{y^\Delta} \ ,
\end{equation}
where $y$ is the perpendicular distance to the the boundary and $\vec{x}$ denotes the remaining translationally invariant coordinates. We immediately see that the distance to the boundary corresponds to the cosmological time and that the temporal `boundary' lies in the infinite future. Two-point functions between a boundary localized operator $\mathcal{O}_3(\vec{x},0)$ and an arbitrarily located operator $\mathcal{O}_4(\vec{x}',y)$ are likewise fully determined
\begin{equation}
	\langle \mathcal{O}_3(\vec{x},0) \mathcal{O}_4(\vec{x}',y)\rangle \sim  \frac{1}{y^{\Delta_4 - \Delta_3}[y^2 + (\vec{x}-\vec{x}')^2]^{\Delta_3}} \ .
\end{equation}
However, correlation functions between two boundary-delocalized operators $\mathcal{O}_4(\vec{x},y)$ and $\mathcal{O}'_4(\vec{x}',y')$ depend upon an unknown function of the conformal invariant $\xi = (\vec{x} - \vec{x}')^2/(4yy')$
\begin{equation}
	\langle\mathcal{O}_4(\vec{x},y)\mathcal{O}_4'(\vec{x}',y')\rangle = \frac{1}{y^{\Delta_4}y'^{\Delta_4'}}f(\xi) \ .
\end{equation}
The form of the function $f(\xi)$ is constrained by the so-called boundary operator product expansion explained in \cite{McAvity:1995zd}. The predicted power spectrum at lowest order in derivatives for a field of weight $\Delta$ field in the pseudo-conformal universe is \cite{Hinterbichler:2011qk}
\begin{equation}
	\langle\mathcal{O}(\vec{k},t)\mathcal{O}(-\vec{k},t)\rangle \sim \frac{1}{k^5 t^{2(\Delta + 1)}} \implies \langle\mathcal{O}(\vec{x},t)\mathcal{O}(0,t)\rangle \sim \frac{\vec{x}^2}{t^{2(\Delta + 1)}} \ ,
\end{equation}
which agrees with the form predicted by \cite{McAvity:1995zd} with $f(\xi) \sim \xi$.

\end{document}